\documentclass[onecolumn]{emulateapj}

\shorttitle{CHARA Array Observations of Vega}
\shortauthors{Aufdenberg et al.}

\begin{document}

\newcommand{\h}{$^{\rm h}$}
\newcommand{\m}{$^{\rm m}$}
\newcommand{\rsun}{\ensuremath{{\rm R}_{\odot}}}
\newcommand{\msun}{\ensuremath{{\rm M}_{\odot}}}
\newcommand{\lsun}{\ensuremath{{\rm L}_{\odot}}}
\newcommand{\teff}{\ensuremath{T_{\rm eff}}}
\newcommand{\kms}{km s$^{-1}$}
\newcommand{\adross}{$\theta_{\rm Ross}$}
\newcommand{\adud}{$\theta_{\rm UD}$}
\newcommand{\adld}{$\theta_{\rm LD}$}

\bibliographystyle{apj}

\title{First Results from the CHARA Array VII: Long-Baseline Interferometric
Measurements of Vega Consistent with a Pole-On, Rapidly Rotating Star}

\author{J. P.~Aufdenberg\altaffilmark{1,2}
A.~M\'{e}rand\altaffilmark{3}, V.~Coud\'{e} du Foresto\altaffilmark{3},
O.~Absil\altaffilmark{4}, E.~ Di Folco\altaffilmark{3}, P.~
Kervella\altaffilmark{3}, S. T.~ Ridgway\altaffilmark{2,3},
D. H.~Berger\altaffilmark{5,6}, T. A.~ten Brummelaar\altaffilmark{6},
H. A.~McAlister\altaffilmark{7}, J.~Sturmann\altaffilmark{6},
L.~Sturmann\altaffilmark{6}, N. H.~Turner\altaffilmark{6}}
\altaffiltext{1}{Michelson Postdoctoral Fellow}
\altaffiltext{2}{National Optical Astronomy Observatory, 950 N. Cherry Ave, Tucson, AZ 85719}
\altaffiltext{3}{LESIA, UMR 8109, Observatoire de Paris-Meudon, 5 place Jules Janssen, 92195 
Meudon Cedex, France}
\altaffiltext{4}{Insitut d'Astrophysique et de G\'{e}ophysique, University of Li\`{e}ge, 
17 All\'{e}e du Six Ao\^{u}t, B-40000  Li\`{e}ge, Belgium}
\altaffiltext{5}{University of Michigan, Department of Astronomy, 500 Church St, 917 Dennison Bldg., Ann Arbor, MI 48109-1042}
\altaffiltext{6}{The CHARA Array, Mount Wilson Observatory, Mount Wilson, CA 91023}
\altaffiltext{7}{Center for High Angular Resolution Astronomy, Department of Physics and Astronomy,
Georgia State University, P.O. Box 3969, Atlanta, GA 30302-3969}

\begin{abstract}
We have obtained high-precision interferometric measurements of Vega
with the CHARA Array and FLUOR beam combiner in the K' band at
projected baselines between 103\,m and 273\,m.  The measured
visibility amplitudes beyond the first lobe are significantly weaker
than expected for a slowly rotating star characterized by a single
effective temperature and surface gravity.  Our measurements, when
compared to synthetic visibilities and synthetic spectrophotometry
from a Roche-von Zeipel gravity-darkened model atmosphere, provide
strong evidence for the model of Vega as a rapidly rotating star
viewed very nearly pole-on. Our model of Vega's projected surface
consists of two-dimensional intensity maps constructed from a library
of model atmospheres which follow pole-to-equator gradients of
effective temperature and surface gravity over the rotationally
distorted stellar surface.  Our best fitting model, in good agreement
with both our interferometric data and archival spectrophotometric
data, indicates that Vega is rotating at $\sim$91\% of its angular
break-up rate with an equatorial velocity of 275 \kms.  Together with
the measured $v\,\sin\,i$, this velocity yields an inclination for the
rotation axis of 5\degr.  For this model the pole-to-equator effective
temperature difference is 2250 K, a value much larger than previously
derived from spectral line analyses.  A polar effective temperature of
10150 K is derived from a fit to ultraviolet and optical
spectrophotometry.  The synthetic and observed spectral energy
distributions are in reasonable agreement longward of 140 nm where
they agree to 5\% or better.  Shortward of 140 nm, the model is up to
10 times brighter than observed.  The far-UV flux discrepancy suggests
a breakdown of von Zeipel's \teff$\propto g^{1/4}$ relation.  The
derived equatorial \teff\ of 7900 K indicates Vega's equatorial
atmosphere may be convective and provides a possible explanation for
the discrepancy.  The model has a luminosity of $\sim$37 \lsun, a
value 35\% lower than Vega's apparent luminosity based on its
bolometric flux and parallax, assuming a slowly rotating star.  The
model luminosity is consistent with the mean absolute magnitude of A0V
stars from the $W({\rm H}\gamma)-M_V$ calibration.  Our model predicts
the spectral energy distribution of Vega as viewed from its equatorial
plane; a model which may be employed in radiative models for the
surrounding debris disk.
\end{abstract}
\keywords{methods: numerical --- stars: atmospheres ---
stars: fundamental parameters (radii, temperature) --- stars: rotation ---
stars:individual (Vega) --- techniques: interferometric}

\section{Introduction}
The bright star Vega ($\alpha$ Lyr, HR 7001, HD
172167, A0 V) has been a photometric standard for nearly 150 years.
\cite{hearnshaw_photometry} describes Ludwig Seidel's visual
comparative photometer measurements, beginning 1857, of 208 stars
reduced to Vega as the primary standard.  Today, precise absolute
spectrophotometric observations of Vega are available from the
far-ultraviolet to the infrared \cite[]{bg04_vega}.  The first signs
that Vega may be anomalously luminous appear in the 1960s after the
calibration of the H$\gamma$ equivalent width to absolute visual
magnitude ($W({\rm H}\gamma) - M_V$) relationship \cite[]{petrie64}.
\cite{MW85} confirmed Petrie's findings using better spectra and
showed that Vega's $M_V$ is 0.5 magnitudes brighter than the mean A0 V
star based on nearby star clusters.  \cite{petrie64} suggested the
anomalous luminosity may indicate that Vega is a binary, however the
Intensity Interferometer measurements by \cite{hb67} found no evidence
for a close, bright companion, a result later confirmed by speckle
observations \cite[]{mcalister_iau_85}.  A faint companion cannot be
ruled out \cite[]{vega_disk06}, however the presence of such a
companion would not solve the luminosity discrepancy.  \cite{hb67}
also noted, based on their angular diameter measurements, that Vega's
radius is 70\% larger than that of Sirius.  Recent precise
interferometric measurements show Vega's radius \cite[R =
2.73$\pm$0.01 \rsun,][]{ciardi01} to be 60\% larger than that of
Sirius \cite[R = 1.711$\pm$0.013 \rsun, M =2.12$\pm$0.06
\msun,][]{kervella_sirius_2003}, while the mass-luminosity and
mass-radius relations for Sirius, $L/\lsun=(M/\msun)^{4.3\pm0.2}$,
$R/\rsun = (M/\msun)^{0.715\pm0.035}$, yield a radius for Vega only
$\sim$12\% larger.

Since the work of \cite{vz1924a,vz1924b}, it has been expected that in
order for rapidly rotating stars to achieve both hydrostatic and
radiative equilibrium, these stars' surfaces will exhibit gravity
darkening, a decrease in effective temperature from the pole to the
equator.  In the 1960s and 1970s considerable effort \cite[see
e.g.,][]{collins63,collins1966,hs68,mp70,cs77} was put into the
development of models for the accurate prediction of colors and
spectra from the photospheres of rapidly rotating stars.  These early
models showed that in the special case where one views these stars
pole on, they will appear more luminous than non-rotating stars, yet
have very nearly the same colors and spectrum.  The connection between
Vega's anomalous properties and the predictions of rapidly rotating model
atmospheres was made by \cite{gray_1985,gray_vega88} who noted that
Vega must be nearly pole-on and rotating at 90\% of its angular
breakup rate to account for its excessive apparent luminosity.
\cite{gray_vega88} also noted that Vega's apparent luminosity is
inconsistant with its measured Str\"{o}mgren color indices which
match that of a dwarf, while the apparent luminosity suggests an
evolved subgiant.

Another anomalous aspect of Vega is the flat-bottom shaped appearance
of many of its weak metal lines observed at high spectral resolution
and very high signal-to-noise ($>$ 2000) \cite[]{gulliver91}.  The
modeling by \cite{elste92} showed that such flat-bottomed or
trapazoidal shaped profiles could result from a strong
center-to-limb variation in the equivalent width of a line coupled
with a latitudinal temperature gradient on the surface of the star.
Soon after, \cite{gha94} modeled these unusual line profiles together
with Vega's spectral energy distribution (SED) and found a nearly
pole-on ($i$=5.5\degr), rapidly rotating ($V_{\rm eq}$ = 245 \kms)
model to be a good match to these data.

Since the detection in the infrared of Vega's debris disk
\cite[]{aumann84}, much of the attention paid to Vega has been focused in
this regard \cite[see e.g.,][]{su05}.  However, not only has Vega's
disk been spatially resolved, so too has its photosphere, first by
\cite{hb67}, though attempts to measure Vega's angular diameter go
back to Galileo \cite[]{hughes2001}. Recent interferometric
measurements of Vega show nothing significantly out of the ordinary
when compared to a standard models for a slowly-rotating A0 V star
\cite[][$v\,\sin\,i$ = 21.9$\pm$0.2 \kms]{hga04}.
Specifically, uniform disk fits to data obtained in the first lobe of Vega's
visibility curve, from the Mark III interferometer \cite[]{m3_03} at
500 nm and 800 nm and from the Palomar Testbed Interferometer (PTI)
\cite[]{ciardi01} in the K-band, show the expected progression due to
standard wavelength-dependent limb darkening: 3.00$\pm$0.05 mas (500
nm), 3.15$\pm$0.03 mas (800 nm), 3.24$\pm$0.01 (K-band).  In addition,
the first lobe data in the optical from the Navy Prototype Optical
Interferometer (NPOI) yield 3.11$\pm$0.01 mas ($\sim$650nm)
\cite[]{ohishi04}, consistent with this picture.  \cite{ciardi01}
note small residuals in their angular diameter fits that may be due to
Vega's disk.

Triple amplitude data from NPOI in May 2001 \cite[]{ohishi04} sample
the second lobe of Vega's visibility curve where a gravity-darkening
signature should be unambiguous, however these data show the signature
of limb darkening expected for a non-rotating star, as predicted by
{\tt ATLAS9} limb-darkening models \cite[]{van_hamme93}. Most
recently, a preliminary analysis of second lobe NPOI data from October
2003 \cite[]{peterson_vega04} indicate that Vega is indeed strongly
gravity darkened, a result inconsistent with
\cite{ohishi04}. \cite{peterson_altair06} note that the NPOI Vega data
are difficult to analyze due to detector nonlinearities for such a
bright star.  \cite{peterson_altair06} do see a strong interferometric
signal for gravity darkening from the rapid rotator Altair with an
angular break-up rate 90\% of critical.  Since a similar rotation rate
is expected for Vega on the basis of its apparently high luminosity
\cite[]{gray_vega88,gha94}, a strong gravity darkening is expected for
Vega as well.

There is clearly a need for additional high spatial resolution
observations of Vega's photosphere to confirm the hypothesis of
\cite{gray_vega88}, confirm the 2003 NPOI observations, and test
the theory of von Zeipel.  We have employed the long baselines of the
CHARA Array \cite[]{theo05} on Mount Wilson, together with the
capabilties of the spatially-filtered Fiber Linked Unit for Optical
Recombination \cite[FLUOR,][]{FLUORatCHARA}, as a means to probe the
second lobe of Vega's visibility curve at high precision and accuracy
in the K-band.  Our Vega campaign, part of the commissioning science
\cite[]{chara_regulus05,delta_cep05,alfcep06} for the CHARA Array, obtained
visibility data on baselines between 103\, m and 273\, m which clearly
show the signature of a strongly gravity darkened, pole-on, rapidly
rotating star.  In this paper we present these data and a detailed
modeling effort to fit both our inteferometric data and the archival data
of Vega's spectral energy distribution.

We introduce our observations in \S\ref{observations}.  Sections
\S\ref{1D_models}, \S\ref{roche_model}, and \S\ref{2-D_fitting}
describe the construction and fitting of one- and two-dimensional
synthetic brightness distributions to our interferometic data and
archival spectrophotometry. A discussion of our results follows in
\S\ref{discussion}. We conclude with a summary in \S\ref{summary}.

\section{Observations}
\label{observations}
\begin{figure*}
\includegraphics[scale=0.6]{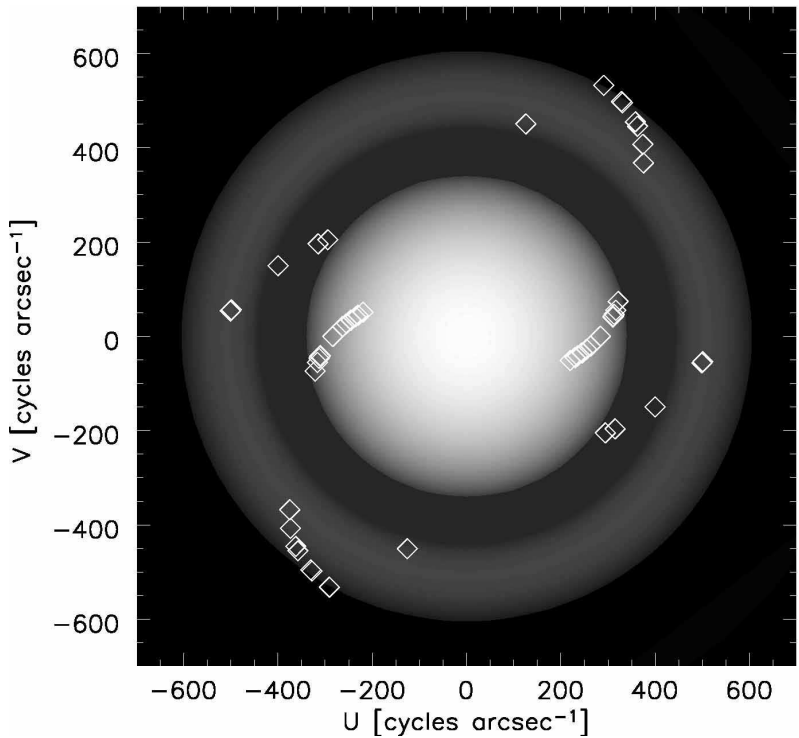}
\caption{The sampling of the $(u,v)$-plane for the CHARA/FLUOR Vega
data set.  The diamonds represent the monochromatic sampling at 2.0
\micron\ within the K' band.  In the K' band, the CHARA baselines E2-W2,
E2-W1, E2-S1, and W2-S1 sample the lower first lobe, first null, and
second lobe of Vega's visibility curve.  Two-telescope observations
have a 180\degr\ ambiguity in the position angle, therefore we plot
two coordinates, $(u,v)$ and $(-u,-v)$, for each of the 25 data
points.  These $(u,v)$ points overlay a model for Vega's
two-dimensional monochromatic Fourier appearance.  This squared
visibility model is a Fast Fourier Transform (displayed with a
logarithmic stretch) of a synthetic intensity map of Vega in the plane
of the sky (see Figure~\ref{fig:surface_map}).}
\label{fig:uv_sampling}
\end{figure*}

Our interferometric measurements were obtained using the Center for
High Angular Resolution Astronomy (CHARA) Array in the infrared
K' band (1.94\micron\ to 2.34\micron) with FLUOR.  Our observations
were obtained during 6 nights in the late spring of 2005 using four
telescope pairs, E2-W2, S1-W2, E2-W1, and S1-E2 with maximum baselines of
156, 211, 251, and 279\,m, respectively.  The FLUOR Data Reduction
Software \cite[]{ksc04,FLUOR_THEORY} was used to extract the squared
modulus of the coherence factor between the two independent telescope
aperatures.  We obtained 25 calibrated observations of Vega which are
summarized in Table~\ref{tab:fluor_data}. The ($u,v$)-plane sampling
is shown in Figure~\ref{fig:uv_sampling}.

\begin{deluxetable*}{rrcrrrrrrl}
\tablecolumns{10} 
\tablecaption{CHARA/FLUOR K'-band Vega Measurements}
\tablewidth{0pt}
\tablehead{ 
\colhead{}
&\colhead{}
&\colhead{}
&\colhead{}
&\colhead{}
&\colhead{Projected} 
&\colhead{Position} 
&\colhead{}
&\colhead{}
&\colhead{}\\
\colhead{}
&\colhead{}
&\colhead{Telescope}
&\colhead{$u$}
&\colhead{$v$}
&\colhead{Baseline}
&\colhead{Angle}
&\colhead{$V^2$}
&\colhead{$\sigma V^2_{\rm total}$}
&\colhead{Calibration Star(s)}\\
\colhead{No.}
&\colhead{Julian Date}
&\colhead{Pair}
&\colhead{(meters)}
&\colhead{(meters)}
&\colhead{(meters)}
&\colhead{(degrees)}
&\colhead{$\times$ 100}
&\colhead{$\times$ 100}
&\colhead{HD Number}
}
\startdata 
1  &2453511.261 &E2-W2 & $-$98.941 &    23.114  &101.606    & $-$76.85         &   21.1531   & 0.8846  & 176527            \\ 
2  &2453511.313 &E2-W2 &$-$127.859 &  $-$0.092  &127.859    &    89.95         &    6.2229   & 0.2019  & 176527, 173780 \\ 
3  &2453511.347 &E2-W2 &$-$139.876 & $-$18.250  &141.062    &    82.56         &    2.6256   & 0.0742  & 173780            \\   
4  &2453511.374 &E2-W2 &$-$144.773 & $-$33.322  &148.558    &    77.03         &    1.3567   & 0.0417  & 173780            \\   
5  &2453512.266 &E2-W2 &$-$103.834 &    20.146  &105.770    & $-$79.02         &   18.2301   & 0.1976  & 159501            \\ 
6  &2453512.269 &E2-W2 &$-$106.062 &    18.698  &107.698    & $-$80.00         &   16.7627   & 0.1710  & 159501            \\  
7  &2453512.277 &E2-W2 &$-$110.513 &    15.601  &111.609    & $-$81.96         &   14.4223   & 0.1493  & 159501            \\   
8  &2453512.284 &E2-W2 &$-$114.716 &    12.396  &115.384    & $-$83.83         &   12.2229   & 0.1336  & 159501            \\   
9  &2453512.291 &E2-W2 &$-$118.435 &     9.291  &118.799    & $-$85.51         &   10.3873   & 0.1168  & 159501            \\   
10 &2453512.345 &E2-W2 &$-$140.179 & $-$18.907  &141.448    &    82.31         &    2.6399   & 0.0741  & 173780            \\   
11 &2453512.349 &E2-W2 &$-$141.068 & $-$20.951  &142.615    &    81.55         &    2.3968   & 0.0676  & 173780            \\   
12 &2453512.356 &E2-W2 &$-$142.577 & $-$24.954  &144.744    &    80.07         &    2.0041   & 0.0591  & 173780            \\   
13 &2453516.258 &E2-W1 &$-$141.950 &    88.392  &167.221    & $-$58.08         &    0.1040   & 0.0059  & 159501            \\  
14 &2453516.343 &E2-W1 &$-$224.986 &    25.325  &226.407    & $-$83.57         &    1.2148   & 0.0521  & 159501, 165683 \\ 
15 &2453517.248 &E2-W1 &$-$132.597 &    92.319  &161.569    & $-$55.15         &    0.2426   & 0.0194  & 159501, 173780 \\ 
16 &2453517.288 &E2-W1 &$-$180.244 &    67.502  &192.469    & $-$69.46         &    0.5913   & 0.0314  & 173780            \\  
17 &2453517.342 &E2-W1 &$-$225.788 &    24.193  &227.080    & $-$83.88         &    1.1066   & 0.0670  & 173780            \\  
18 &2453519.225 &E2-S1 &$-$169.006 &$-$165.745  &236.716    &    45.55         &    1.1361   & 0.0414  & 159501            \\  
19 &2453519.252 &E2-S1 &$-$168.472 &$-$183.482  &249.095    &    42.55         &    0.9120   & 0.0344  & 159501            \\  
20 &2453519.285 &E2-S1 &$-$161.265 &$-$205.029  &260.851    &    38.18         &    0.6047   & 0.0259  & 159501            \\ 
21 &2453519.316 &E2-S1 &$-$147.913 &$-$224.292  &268.673    &    33.40         &    0.5079   & 0.0238  & 159501            \\  
22 &2453522.270 &E2-S1 &$-$163.306 &$-$200.735  &258.773    &    39.13         &    0.5911   & 0.0427  & 159501            \\  
23 &2453522.306 &E2-S1 &$-$148.868 &$-$223.205  &268.295    &    33.70         &    0.4518   & 0.0241  & 159501            \\  
24 &2453522.336 &E2-S1 &$-$131.105 &$-$239.777  &273.279    &    28.67         &    0.3788   & 0.0199  & 159501            \\  
25 &2453538.206 &W2-S1 &    56.624 &   202.948  &210.699    &    15.59         &    0.9303   & 0.0682  & 162211            \\ 
\enddata
\label{tab:fluor_data}
\end{deluxetable*}

The calibrator stars were chosen from the catalogue of
\cite{antoine_cals_2005}.  The CHARA Array's tip-tilt adaptive optics
system operates at visual wavelengths.  Vega is sufficiently bright
that it was necessary to reduce the gain on the tip-tilt detector
system while observing Vega and return the gain to the nominal setting
for the fainter calibrator stars.  Calibrators chosen for this work
are all K giants: HD 159501 (K1 III), HD 165683 (K0 III), HD 173780
(K3 III), HD 176567 (K2 III), and HD 162211 (K2 III).  The spectral
type difference between the calibrators and Vega does not
significantly influence the final squared visibility estimate.  The
interferometric transfer function of CHARA/FLUOR was estimated by
observing a calibration star before and after each Vega data point.
In some cases a different calibrator was used on either side of the
Vega data point (see Table~\ref{tab:fluor_data}).  The inteferometric
efficiency of CHARA/FLUOR was consistent between all calibrators and
stable over each night at $\sim$85\%.


\section{One-Dimensional Model Fits}
\label{1D_models}
Under the initial assumption that Vega's projected photospheric disk
is circularly symmetric in both shape and intensity, we have fit
three models to the CHARA/FLUOR data set: (1) a uniform disk,
where the intensity, assumed to be Planckian $I(\lambda) =
B(\teff = 9550 {\ \rm K} ,\lambda)$, is independent of $\mu$, the
cosine of the angle between the line-of-sight and the surface normal;
(2) an analytic limb-darkening law, $I(\mu,\lambda)=B(\teff = 9550 {\
\rm K},\lambda)\mu^\alpha$; and (3) a {\tt PHOENIX} \cite[]{nextgen2}
model radiation field with stellar parameters (\teff = 9550 K,
$\log(g)$ = 3.95) consistent with the slowly rotating model that
\cite{bg04_vega} show to be a good match to Vega's observed SED.  The
computation of the synthetic squared visibilities from these models
takes into account the bandwidth smearing introduced by the
non-monochromatic FLUOR transmission (see \S~\ref{bws} below).

Figure~\ref{fig:alpha_fit} shows the synthetic squared visibilities
from the three models in comparison with the CHARA/FLUOR data.  The
uniform disk angular diameter we derive is ($\theta_{\rm UD} =
3.209\pm0.003$ mas) is not consistent with \cite{ciardi01},
$\theta_{\rm UD} = 3.24\pm0.01$ mas.  We find this is most likely
because we do not assume a flat spectrum across the K' band filter.
Regardless, this uniform disk model is poor fit ($\chi^2_\nu$ = 38)
because it neglects limb darkening.  The limb darkening expected for a
slowly rotating star should be well predicted by the {\tt PHOENIX}
model, but this model is also a poor fit ($\chi^2_\nu = 20,
\theta_{\rm LD} = 3.259\pm0.002$ mas).  The second lobe data indicate
Vega is significantly more limb darkened compared to this model.  The
non-physical $I(\mu)=\mu^\alpha$ model yields a much better fit
($\chi^2_\nu$ = 1.5) and a significantly larger angular diameter
$\theta_{\rm LD} = 3.345\pm0.006$ ($\alpha = 0.341\pm0.013$), which
suggests the limb-darkening correction in the K' band is $\sim$2.5
times larger (4.2\% vs.\,1.6\%) than expected for a slowly-rotating
Vega.

\begin{figure*}
\includegraphics[scale=0.6,angle=0.0]{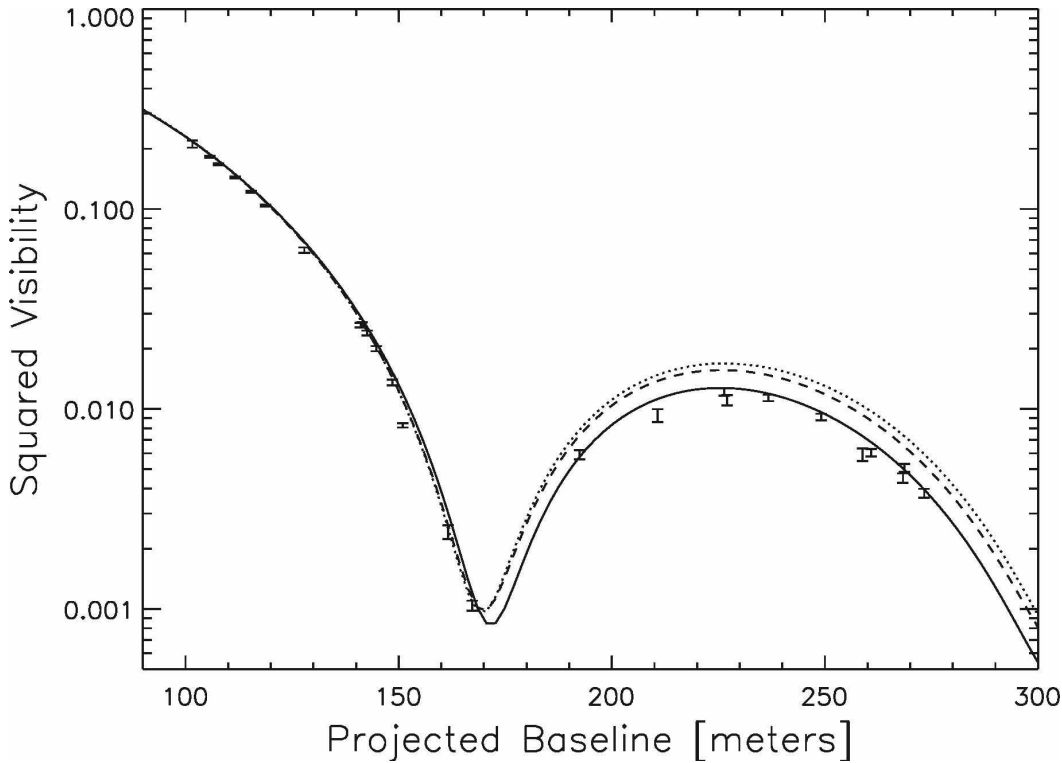}
\caption{Best fit one-dimensional, symmetric
models in comparison with the CHARA/FLUOR data set.  The dotted line is a
bandwidth-smeared uniform disk ($\chi^2_\nu = 38, \theta_{\rm UD} =
3.209\pm0.003$ mas) The dashed line is a bandwidth-smeared {\tt PHOENIX}
model atmosphere with parameters consistent with a slowly rotating
Vega, ($\teff\ = 9550 {\ \rm K}, \log(g) = 3.95, \chi^2_\nu = 20,
\theta_{\rm LD} = 3.259\pm0.002$ mas), and the solid line a
bandwidth-smeared analytic limb-darkening model, $I(\mu) = \mu^\alpha$
($\chi^2_\nu = 1.5, \theta_{\rm LD} = 3.345\pm0.006 {\ \rm mas}, \alpha = 0.341\pm0.013$).
If extended emission in the K' band is present at the 1.3\% level
in the Vega system, these best angular diameters are systematically high
by $\sim$3$\sigma$ (see text).}
\label{fig:alpha_fit}
\end{figure*}

\cite{vega_disk06} report that a small fraction, $f=$1.29$\pm$0.16\%,
of Vega's K' band flux comes from an extended structure, most likely
Vega's debris disk.  In order to gauge the significance of this extra
flux on the photospheric parameters derived above, the synthetic
squared visibilities are reduced by an amount equal to the square of
fraction of light coming from the debris disk.  At long baselines, the
visibility of the debris disk is essentially zero such that:
\begin{eqnarray}
\label{eqn:vobs} 
V^2_{obs} &=& \biggl[(1-f)V_{\rm photosphere} + fV_{\rm disk}\biggr]^2\\
            &\approx& 0.974 V_{\rm photosphere}^2 \nonumber
\end{eqnarray}
The revised fits to $V_{\rm photosphere}^2$ are $\theta_{\rm UD} =
3.198\pm0.003$ ($\chi^2_\nu$ = 38) for the uniform disk, $\theta_{\rm
  LD} = 3.247\pm0.002$ ($\chi^2_\nu$ = 19) for the {\tt PHOENIX}
model, and $\theta_{\rm LD} = 3.329\pm0.006$ ($\alpha =0.328\pm0.013$,
$\chi^2_\nu$=1.4), for the $I(\mu)=\mu^\alpha$ model. The effect of
removing the extended emission is to reduce the best fit angular diameter
for all three models by $\sim$3$\sigma$; the correction for extended
emission is therefore significant.

\section{Two-Dimensional Model Construction}
\label{roche_model}
In order to physically interpret the strong limb darkening we find for
Vega, we have adapted a computer program written by S. Cranmer
(private communication, 2002) from \cite{co95} which computes the
effective temperature and surface gravity on the surface of a
rotationally distorted star, specifically a star with an infinitely
concentrated central mass under uniform angular rotation, a Roche-von
Zeipel model.  This azimuthally symmetric model is parameterized as a
function of the colatitude given the mass, polar radius, luminosity,
and fraction of the angular break-up rate.

Each two-dimensional intensity map is characterized by five variables:
$\theta_{\rm equ}$, the angular size of the equator, equivalent to the
angular size as viewed exactly pole-on; $\omega=\Omega/\Omega_{\rm
crit}$, the fraction of the critical angular break-up rate; $T^{\rm
pole}_{\rm eff}$, the effective temperature at the pole; $\log(g)_{\rm
pole}$, the effective surface gravity at the pole; and $\psi$, the
position angle of the pole on the sky measured east from north.

Given these input parameters, along with the measured trigonometric
parallax $\pi_{\rm hip}$ = 128.93$\pm$0.55 mas \cite[]{hipparcos}, and the
observed projected rotation velocity, $v\,\sin\,i$ = 21.9$\pm$0.2
\kms\ \cite[]{hga04}, the parameterization of the intensity maps
begins with
\begin{equation}
R_{\rm equ} = 107.48\frac{\theta_{\rm equ}}{\pi_{\rm hip}}
\label{eqn:requ} 
\end{equation}
the equatorial radius in solar units with both $\theta_{\rm equ}$ and
$\pi_{\rm hip}$ in milliarcseconds.
It follows from a Roche model \cite[equation 26]{co95} that the corresponding polar
radius is
\begin{equation}
R_{\rm pole} = \frac{\omega\, R_{\rm equ}}
{3\,\cos\biggl[\frac{\displaystyle \pi+\cos^{-1}(\omega)}{\displaystyle 3}\biggr]}
\label{eqn:rpole} 
\end{equation}
and the stellar mass is
\begin{equation}
M = \frac{g_{\rm pole}\,R_{\rm pole}^2}{G} 
\label{eqn:mass}
\end{equation}
where $G$ is the universal gravitational constant.
The luminosity is then, 
\begin{equation}
L = \frac{\sigma\Sigma(T_{\rm eff}^{\rm pole})^4}{g_{\rm pole}}
\label{eqn:lum}
\end{equation}
where $\sigma$ is the Stefan-Boltzman constant and $\Sigma$ is the
surface-weighted gravity $\Sigma$ \cite[equations 31 and 32]{co95},
expressed as a power series expansion in $\omega$,
\begin{eqnarray}
&\Sigma \approx\  4\pi GM\biggl[1.0 - 0.19696\,\omega^2 - 
0.094292\,\omega^4 + 0.33812\,\omega^6& \nonumber \\ 
& - 1.30660\,\omega^8 + 1.8286\,\omega^{10} - 0.92714\,\omega^{12}\biggr]&
\label{eqn:sigma}
\end{eqnarray}

The ratio of the luminosity to $\Sigma$ provides the proportional
factor between the effective temperature and gravity for von
Zeipel's radiative law for all colatitudes $\vartheta$:
\begin{equation}
T_{\rm eff}(\vartheta) = \biggl[\frac{L}{\sigma\Sigma}g(\vartheta)\biggr]^{\beta} = 
T_{\rm eff}^{\rm pole} \biggl[\frac{g(\vartheta)}{g_{\rm pole}}\biggr]^{\beta} 
\label{eqn:vz}
\end{equation}
where the gravity darkening parameter, $\beta$, takes the value 
0.25 in the purely radiative case (no convection). The effective temperature
difference between the pole and equator, $\Delta\teff$, may be expressed 
in terms of $T_{\rm eff}^{\rm pole}$ and $\omega$:
\begin{equation}
\Delta\teff = T_{\rm eff}^{\rm pole} - T_{\rm eff}^{\rm equ} = 
T_{\rm eff}^{\rm pole}
\Biggl( 1 - \biggl[\frac{\omega^2}{\eta^2}-\frac{8}{27}\eta\omega\biggr]^\beta\Biggr) 
\label{eqn:deltateff}
\end{equation} 
where,
\begin{displaymath}
\eta = 3\,\cos\biggl[\frac{\displaystyle \pi+\cos^{-1}(\omega)}{\displaystyle 3}\biggr]
\end{displaymath} 

The effective gravity as a function of $\vartheta$ is given by
\begin{eqnarray}
g(\vartheta) &=&  \biggl[ g_r(\vartheta)^2 + g_\vartheta(\vartheta)^2 \biggr]^{1/2}   \\
g_r(\vartheta) &=& \frac{-GM}{R(\vartheta)^2} +R(\vartheta)(\Omega \sin\vartheta)^2  \\
g_\vartheta(\vartheta) &=& R (\vartheta)\Omega^2 \sin\vartheta\cos\vartheta 
\end{eqnarray}
where $g_r$ and $g_\vartheta$ are the radial and colatitudinal components of
the gravity field.  The colatitudinal dependence of the radius is given by
\begin{equation}
R(\vartheta) = 3\frac{R_{\rm pole}}{\omega\sin\vartheta} 
\cos \biggl[\frac{\pi + \cos^{-1}(\omega\sin\vartheta)}{3}\biggr]\quad (\omega > 0)
\end{equation}
and angular rotation rate is related to the critical angular rotation
rate \footnote{There is a typographical error in equation (5) of
\cite{collins63} which is not in the paper's erratum
\cite[]{collins64_erratum}: $\omega^2_{c} = \frac{GM}{R_e}$ should be
$\omega^2_{c} = \frac{GM}{R_e^3}$, where $\omega_c$ the critical angular
rate, and $R_e$ is the equatorial radius at the critical rate.} by
\begin{equation}
\Omega = \omega \Omega_{\rm crit} = \omega \biggl[\frac{8}{27}\frac{GM}{R^3_{\rm pole}}\biggr]^{1/2}
\label{eqn:Omega}
\end{equation}

At the critical rate ($\omega = 1$), $R_{\rm equ} = 1.5 R_{\rm pole}$.  
The inclination follows from
\begin{equation}
i = \sin^{-1} \biggl[\frac{v\sin i}{V_{\rm equ}} \biggr]
\label{eqn:i}
\end{equation}
where the equatorial velocity is 
\begin{equation}
V_{\rm equ}  = R_{\rm equ}\, \Omega.
\label{eqn:vequ}
\end{equation}

\subsection{Building the Intensity Maps}
\label{build}
For each wavelength, $\lambda$ (185 total wavelength points:
1.9\micron\ to 2.4\micron\ in 0.005\micron\ steps, with additional
points for \ion{H}{1} and \ion{He}{1} profiles calculated in non-LTE), an
intensity map is computed as follows: \teff$(\vartheta)$ and
$\log(g(\vartheta)$ are evaluated at 90 $\vartheta$ points from
0\degr\ to 90\degr\ + $i$.  At each $\vartheta$ there are 1024
longitude $\varphi$ points from 0\degr\ to 360\degr\ to finely sample
the perimeter of the nearly pole-on view.  For Vega's nearly pole-on
orientation, the relatively high resolution in $\varphi$ reduces
numerical aliasing when the brightness map is later interpolated onto
a uniformly gridded rectangular array as described below.

Each set of spherical coordinates ($R(\vartheta)$, $\vartheta$,
$\varphi$) is first transformed to rectangular ($x$, $y$, $z$)
coordinates with the Interactive Data Language (IDL) routine {\tt
POLEREC3D}\footnote{The coordinate transformation routines used here
are from the JHU/APL/S1R IDL library of the Space Oceanography Group
of the Applied Physics Laboratory of The Johns Hopkins University.}.
Next, the $z$-axis of the coordinate system is rotated away from the
observer by an angle equal to the inclination $i$ (using the IDL
routine {\tt ROT\_3D}) and then the $x$-$y$ plane is rotated by an angle
equal to $\psi$, the position angle (east of north) of the pole on the
sky (using the IDL routine {\tt ROTATE\_XY}).

At each point in the map, the cosine of the angle between the observer's
line-of-sight and the local surface normal is
\begin{eqnarray}
\mu(x,y) = \mu(\vartheta,\varphi,i) = &&\nonumber \\
\frac{1}{g(\vartheta)}\biggl\{-g_r(\vartheta)[\sin\vartheta\sin i \cos\varphi + 
\cos\vartheta\cos i]&& \nonumber \\ 
-g_\vartheta(\vartheta)[\sin i \cos\varphi\cos\vartheta - \sin\vartheta\cos i]\biggr\}.&&
\end{eqnarray}
The intensity at each point $(x,y)$ is interpolated from a grid of 170
spherical, hydrostatic {\tt PHOENIX} (version 13.11.00B) stellar atmosphere models
\cite[]{nextgen2} spanning 6500 K to 10500 K in \teff\
and 3.25 to 4.15 in $\log(g)$:
\begin{eqnarray*}
    T_j           = 6500 + 250\cdot j \quad {\rm K}&\quad &j=\{0,1,\ldots,16\} \\
    \log(g_l)     = 3.25 + 0.1\cdot l              &\quad &l=\{0,1,\ldots,9\}.
\end{eqnarray*}
Four radiation fields, $I(\lambda, \mu)$ evaluated at 64 angles by
{\tt PHOENIX}, are selected from the model grid to bracket the local
effective temperature and gravity values on the grid square,
\begin{eqnarray*}
T_j  < &T_{\rm eff}(\vartheta) &< T_{j+1} \\
g_l  < &g(\vartheta)  &< g_{l+1}.
\end{eqnarray*}
The intensity vectors $I_\lambda(\mu)$ are linearly interpolated (in the log)
at $\mu(x,y)$ around the grid square,

\begin{eqnarray*}
I_\lambda^{00} &=& I_\lambda(T_j,g_l,\mu(x,y)) \\
I_\lambda^{10} &=& I_\lambda(T_{j+1},g_l,\mu(x,y)) \\
I_\lambda^{11} &=& I_\lambda(T_{j+1},g_{l+1},\mu(x,y)) \\
I_\lambda^{01} &=& I_\lambda(T_j,g_{l+1},\mu(x,y)).
\end{eqnarray*}
Next, the intensity is bilinearly interpolated at the
local $T_{\rm eff}$ and $\log(g)$ for each $(x,y)$ position in the map:
\begin{eqnarray}
I_\lambda(x,y) &=&  I_\lambda[T_{\rm eff}(x,y),\,g(x,y),\, \mu(x,y)]\nonumber\\
               &=& (1-a)(1-b)\, I_\lambda^{00} + a(1-b)\, I_\lambda^{10} \nonumber\\
               &&  +ab\, I_\lambda^{11} + (1-a)b\, I_\lambda^{01} 
\end{eqnarray}
where
\begin{eqnarray*}
a              &=& (T_{\rm eff}(x,y) - T_j)/(T_{j+1} - T_j)\\
b              &=& (g(x,y)- g_l)(g_{l+1}-g_l)
\end{eqnarray*}

Finally, a Delaunay triangulation is computed (using the IDL routine
{\tt TRIGRID}) to regrid the intensity map $I_\lambda(x,y)$, originally gridded
in $\vartheta$ and $\varphi$, onto a regular 512x512 grid of points
in $x$ and $y$.  The coordinates $x$ and $y$ have the units of
milliarcseconds and correspond to offsets in right ascension and
declination on the sky ($\Delta\alpha$,$\Delta\delta$) relative to the
origin, the subsolar point. 


\subsection{Synthetic Squared Visibility Computation}
\label{synvis}
Due to the lack of symmetry in the synthetic intensity maps, we
evaluate a set of discrete 2-D Fourier transforms in order to generate
a set of synthetic squared visibilities comparable to the CHARA/FLUOR
observations.  The first step is to compute the discrete Fourier
transform for each wavelength at each of the spatial frequency
coordinates $(u,v)$ corresponding to the projected baseline and
orientation of each data point (see Table~\ref{tab:fluor_data}).  The
mean $(u,v)$ coordinates for each data point, in units of meters, are
converted to the corresponding spatial frequency coordinates
$(u_k,v_k)$ in units of cycles per arcsecond for each wavelength
$\lambda_k$. The Fourier transform,
\begin{equation}
V^2_\lambda (u,v) = \Biggl[\int_{-\infty}^{\infty}\int_{-\infty}^{\infty} 
S_\lambda I_\lambda(x,y) e^{i2\pi(u\,x + v\,y)}\, \mathrm{d}x\, \mathrm{d}y \Biggr]^2
\end{equation}
is approximated by the integration rule of Gaussian quadrature \cite[e.g.,][]{stroud_secrest66,numrec}

\begin{eqnarray}
V_k^2 (u_k,v_k) &\approx  &\Biggl[ \sum_{i=1}^N A_i \sum_{j=1}^N A_j  S_k I_k(x_i,y_j) \cos(2\pi(u_k x_i + v_k y_j)) \Biggr]^2 \nonumber\\
&+ &\Biggr[ \sum_{i=1}^N A_i \sum_{j=1}^N A_j  S_k I_k(x_i,y_j) \sin(2\pi(u_k x_i + v_k y_j)) \Biggr]^2
\label{eqn:quadrature}
\end{eqnarray}
where $S_k$ is the wavelength discretized value of the instrument
sensitivity curve $S_\lambda$, and $A_i$, $A_j$ and $x_i$, $y_j$ are
the weights and nodes of the quadrature, respectively.  For our square
grid, the $x-$ and $y-$coordinate nodes and weights are indentical.
The 2-D Gaussian quadrature is performed with a version of the IDL
routine {\tt INT\_2D} modified to use an arbitrarily high number
of nodes.  The intensity at wavelength $k$, $I_k(x,y)$, is interpolated
at $(x_i,y_j)$ from the regular $512\times512$ spacing to the
quadrature node points using the IDL routine {\tt INTERPOLATE} which
uses a cubic convolution interpolation method employing 16 neighboring
points.  The synthetic squared visibility is normalized to unity at
zero spatial frequency by:
\begin{equation}
V_k^2 (0,0) \approx \Biggl[\sum_{i=1}^N A_i \sum_{j=1}^N A_j S_k I_k(x_i,y_j)\Biggr]^2.
\label{eqn:norm_quadrature}
\end{equation}

We find $N=512$ provides the degree of numerical accuracy sufficient in the case of a 2-D
uniform disk (right circular cylinder) to yield $V^2$ values in
agreement with the analytic result,
\begin{equation}
V_k^2 (u_k,v_k) = \Biggl[2 J_1(\pi\theta\sqrt{u_k^2 + v_k^2})/(\pi\theta \sqrt{u_k^2 + v_k^2})\Biggr]^2
\label{eqn:ud}
\end{equation}
(where $J_1$ is the first order Bessel function of the first kind,
$\theta$ is the angular diameter of the uniform disk and $B$ is the
projected baseline), to better than 1\% for $V^2 > 10^{-3}$. We use
the IDL function {\tt BESELJ} for our $J_1$ computations.  For $V^2
\lesssim 10^{-4}$, near the monochromatic first and second zeros, the
numerical accuracy of the quadrature deteriorates to 10\% or worse.
The bandwidth-smeared $V^2$ minimum is at $\sim 10^{-3}$, so we find
this quadrature method yields squared visibilities which are sufficiently
accurate for our task, however observations with an even larger
dynamic range \cite[]{ps05} will require more accurate methods.

\begin{figure*}
\includegraphics[scale=0.6,angle=0.0]{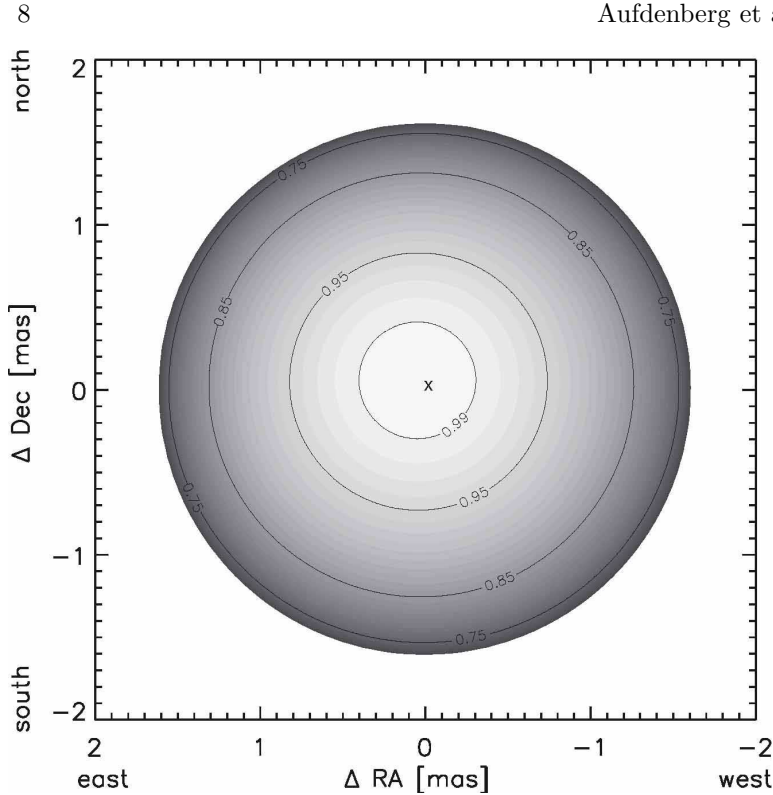}
\caption{A synthetic brightness map (linear stretch) of Vega for our best fitting
parameters: $\omega=0.91$, $\theta_{\rm equ} = 3.329$ mas, $T^{\rm
pole}_{\rm eff} =$ 10150 K, $\log(g)_{\rm pole} =$ 4.10. 
For this model, Vega's pole is inclined 5\degr\ toward a position angle of 40\degr\ 
and 'x' marks the subsolar point.  The labeled intensity contours
are relative to the maximum intensity in the map.}
\label{fig:surface_map}
\end{figure*}

To test the 2-D Gaussian quadrature method in the case where no
analytic solution is available, we computed the 2-D Fast Fourier
Transform (IDL routine {\tt FFT}) of a brightness map (see
Figure~\ref{fig:surface_map}).  First, we compared the results of the
2-D FFT to the analytic uniform disk, equation (\ref{eqn:ud}).  To
reduce aliasing we find it necessary to ``zero pad'' the brightness
map.  With 12-to-1 zero padding (the $512\times512$ brightness map
placed at the center of a larger $6144\times6144$ array of zeros) we
find the 2-D FFT has very similar accuracy to the 512-point Gaussian
quadrature: better than 1\% down to $V^2 \gtrsim 10^{-3}$ inside the
second null.  For the brightness map shown in
Figure~\ref{fig:surface_map}, the 2-D FFT and Gaussian quadrature
methods agree to better than 0.5\% down to $V^2 \gtrsim 10^{-3}$,
inside the second null. We find the computational time required to
evaluate equation (\ref{eqn:quadrature}) at 25 $(u_k,v_k)$ points for
185 wavelengths is $\sim 6$ times faster than the evaluation of the 185
zero-padded 2-D FFTs.

\subsubsection{Bandwidth Smearing} 
\label{bws}
Once we have computed $V^2_k (u_k,v_k)$ for the 185 wavelength
points, we proceed to compute the bandwidth-smeared average
squared visibility $V(B,\lambda_0)^2$,
\begin{equation}
V(B,\lambda_0)^2 = \frac{\int_0^{\infty} V(B,\lambda)^2\, \lambda^2 \, \mathrm{d}\lambda}
{\int_0^{\infty} V(0,\lambda)^2\, \lambda^2 \, \mathrm{d}\lambda}.
\end{equation}
This integral is performed by the IDL routine {\tt INT\_TABULATED}, a 5-point
Newton-Cotes formula.  The $\lambda^2$ term is included so that
the integral is equivalent to an integral over wavenumber (frequency) where

\begin{equation}
\lambda_0^{-1} = \frac{\int_0^{\infty}  \lambda^{-1}\ S(\lambda)\ F_{\lambda}\ \, \mathrm{d}\lambda}
{\int_0^{\infty}  S(\lambda)\ F_{\lambda}\  \, \mathrm{d}\lambda} ,
\end{equation}
is the mean wavenumber.  This simulates the data collection and fringe processing algorithm
used by FLUOR.
In the discretized integrand, $V(B,\lambda_k)^2$ is equivalent to
$V^2_k(u_k,v_k)$ where $B = 206264.8 \lambda_k \sqrt{u_k^2 +v_k^2}$,
with $\lambda_k$ in units of meters and $u_k$ and $v_k$ in units of
cycles per arcsecond.  

\subsection{Synthetic Spectral Energy Distribution Construction}
\label{SED}
To construct synthetic SEDs for Vega from the Roche-von Zeipel model,
170 radiation fields were computed from the same model grid used to
construct the K' band intensity maps.  The wavelength resolution is
0.05 nm from 100 nm to 400 nm and 0.2 nm from 400 nm to
3 \micron\ and 10 nm from 3 \micron\ to 50 \micron.  The higher
resolution in the ultraviolet is needed to sample the strong line
blanketing in this spectral region.
From the resulting grid of radiation fields, intensity maps
are computed (see \ref{build}) and the flux is computed from
\begin{equation}
F_\lambda = \int_0^\pi\int_0^{2\pi}    -\frac{\displaystyle g(\vartheta)}{\displaystyle g_r(\vartheta)}
 I_\lambda(R,\vartheta,\varphi) R(\vartheta)^2 \sin\vartheta \mu (\vartheta,\varphi,i)\,\mathrm{d}\varphi\,\mathrm{d}\vartheta .
\label{eqn:flux}
\end{equation}
This 2-D integral is performed with the IDL routine {\tt
INT\_TABULATED\_2D} (version 1.6) which first constructs a Delaunay
triangulation of points in the $\vartheta\varphi$-plane. For each
triangle in the convex hull (defined as the smallest convex polygon
completely enclosing the points), the volume of the triangular cylinder
formed by six points (the triangle in the plane and three points above
with heights equal to the integrand) is computed and summed.  For
computing the flux from the intensity maps, a coarser sampling in
$\vartheta$ and $\varphi$ (20$\times$40), relative to that needed for
the visibility computations, is sufficient for better than 1\% flux
accuracy.  The numerical accuracy was checked by computing the SED of
a non-rotating star ($\omega$ = 0) and comparing this to a single
effective temperature SED from a 1-D atmosphere.  The interpolation
and integration errors result in a flux deficit of less than 0.7\% at
all wavelengths relative to the 1-D model atmosphere.

\section{Two-Dimensional Model Fitting}
\label{2-D_fitting}
\subsection{Initial Parameter Constraints}
The computation of each intensity map, the Fourier transforms, and the
bandwidth-smearing for each set of input parameters ($\theta_{\rm equ}$,
$\omega$, $T^{\rm pole}_{\rm eff}$, $\log(g)_{\rm pole}$, $\psi$) is
too computationally expensive to compute synthetic squared
visibilities many hundreds of times as part of a gradient-search
method over the vertices of a 5-dimensional hypercube.  Therefore, we
must proceed with targeted trial searches to establish the sensitivity
of $\chi^2_\nu$ to each parameter after first establishing a reasonable
range of values for each parameter.

The parameter $\theta_{\rm equ}$ is a physical angular diameter
related to a uniform disk fit by a scale
factor depending on the degree of gravity and limb darkening, which in
turn depends on the parameters $\omega$, $\log(g)_{\rm pole}$, and
$T^{\rm pole}_{\rm eff}$, in order of importance.  As shown above, a
limb-darkening correction of 4\% is significantly larger than the
$\sim$1.5\% value expected for a normal A0 V star at 2\micron\
\cite[]{dtb00}.  The analytic limb-darkening model fit is sufficiently
good that we take $\theta_{\rm equ}$ = 3.36 mas as a starting value.
This corresponds to $R_{\rm equ}$ = 2.77 \rsun\ from equation
(\ref{eqn:requ}).

Our starting value for $\omega$ is based on the assumption that Vega's
{\it true} luminosity should be similar to that slowly rotating A0 V
stars.  Vega has an apparent luminosity, assuming a single effective
temperature from all viewing angles, of 57 \lsun\ based on its
bolometric flux and the parallax. In the pole-on rapidly rotating
case, we would see Vega in its brightest projection.
According to \cite{MW85} the mean
absolute visual magnitude, $M_V$, is 1.0 for spectral type A0 V.  With
a bolometric correction of $-$0.3, this translates to $L = 37.7$
\lsun.  From equations (\ref{eqn:lum}) and (\ref{eqn:sigma}) we expect
$\omega > 0.8$ in order to account for the luminosity discrepancy assuming a
minimum polar effective temperature of 9550 K, based on the non-rotating
model fits to Vega's SED \cite[]{bg04_vega}.  The range of effective
temperatures and surface gravities for the model atmosphere grid
described in \S\ref{build} sets our upper rotation limit at $\omega
\leq 0.96$.  For $\omega >0.8$,  $\Delta\teff > 1300$ K (see equation \ref{eqn:deltateff}),
thus $T^{\rm pole}_{\rm eff}$ must be greater
than 9550 K to compensate for the pole-to-equator temperature
gradient and to reproduce the observed SED.  
So, given a mean apparent \teff\ of 9550 K, a rough estimate
of $T^{\rm pole}_{\rm eff}$ is 9550 K + $\frac{1}{2} \Delta\teff$ = 10200 K.
We therefore limit the polar effective temperature to the range 10050
K $ < T^{\rm pole}_{\rm eff} < $ 10350 K.

The relationship between $\omega$ and the true luminosity, through
equations (\ref{eqn:lum}), (\ref{eqn:sigma}), and (\ref{eqn:mass}), is
independent of the polar surface gravity, yet we can constrain
$\log(g)_{\rm pole}$ by assuming Vega follows the mass-luminosity
relation we derive for the slowly rotating Sirius,
$L/\lsun=(M/\msun)^{4.3\pm0.2}$.  Here we assume Vega's rapid rotation
has no significant effect on its interior in relation to the
luminosity from nuclear reactions in its core.  Assuming $L=37.7$
\lsun\ from above, the mass-luminosity relation yields $M$=2.3$\pm$0.1
\msun.  As $\omega$ increases, $R_{\rm pole}$ decreases relative to
$R_{\rm equ}$, therefore choosing $M$ = 2.2 \msun\ and $\omega = 0.8$
provides a lower limit of $\log(g)_{\rm pole} = 4.0$.  For lower polar
gravities, Vega's mass will be significantly lower than expected based
on its luminosity, nevertheless we choose a range $\log(g)_{\rm pole}$
values from 3.6 and 4.3 in order to check the effect of the gravity on
our synthetic visibilities and SEDs.

Lastly, the position angle of Vega's pole, $\psi$, should be important
if Vega's inclination is sufficiently high {\it and} its rotation
sufficiently rapid to produce an elliptical projection of the
rotationally distorted photosphere on plane of the sky.  Previous
measurements \cite[]{ohishi04,ciardi01} find no evidence for
ellipticity. Preliminary results from the NPOI three-telescope
observations of \cite{peterson_vega04} suggest an asymmetric
brightness distribution with $\psi$ = 281\degr.

\subsection{CHARA/FLUOR Data: Parameter Grid Search} 
\label{2dgridsearch}
\begin{figure*}
\includegraphics[scale=0.6,angle=0.0]{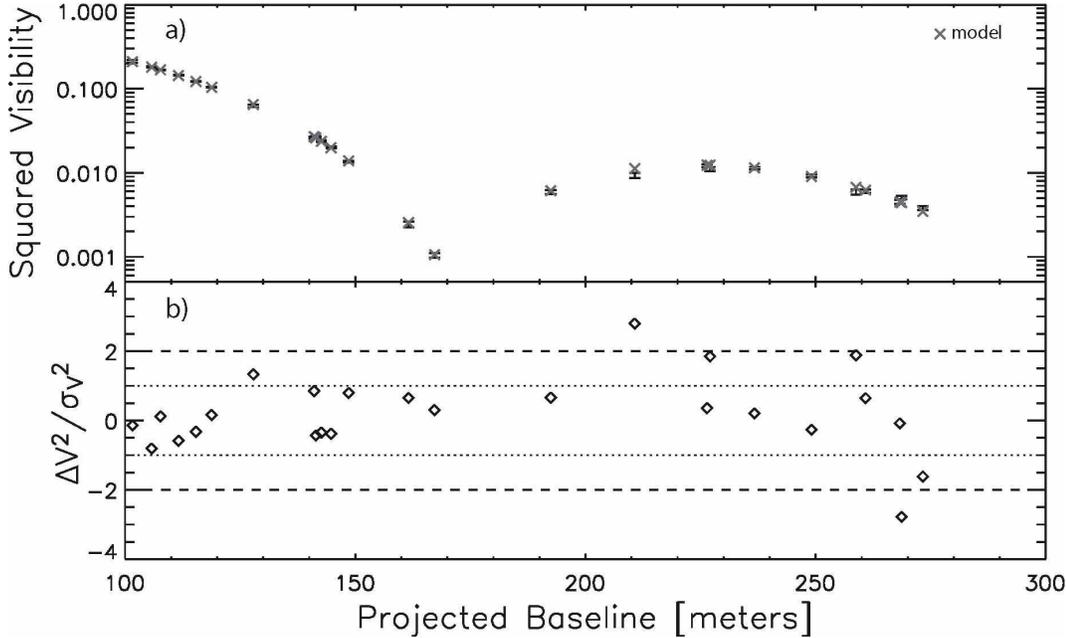}
\caption{a) The CHARA/FLUOR $V^2$ data (error bars) plotted as a
function of projected baseline (for a range of azimuths, see
Table~\ref{tab:fluor_data}) together with the best fitting Roche-von
Zeipel synthetic squared visibilities. Model parameters:
$\omega$=0.91, $\theta_{\rm equ}$ = 3.329 mas, $T^{\rm pole}_{\rm
eff}$ = 10250 K, $\log(g)_{\rm pole}$ = 4.10.  The best fit
$\chi_\nu^2$ = 1.31.  b) Deviations of the best-fit model from
observed squared visibilities.  The dotted and dashed lines indicated
the 1$\sigma$ and 2$\sigma$ deviations.}
\label{fig:v2_bestfit}
\end{figure*}

For the grid search we compute the reduced chi-square $\chi_\nu^2$ for
a set of models defined by $\theta_{\rm equ}$, $\omega$, $T^{\rm
pole}_{\rm eff}$, $\log(g)_{\rm pole}$, and $\psi$, adjusting
$\theta_{\rm equ}$ slightly ($<$0.3\%) to minimize $\chi_\nu^2$ for
each model (see below).  Figure~\ref{fig:pa_vs_omega} shows a
$\chi_\nu^2$ map in the $\omega - \psi$ plane for a range of
$\theta_{\rm equ}$ values with $T^{\rm pole}_{\rm eff}$ = 10250 K.
$\log(g)_{\rm pole}$ = 4.1.  We find a best fit of $\chi_\nu^2$ = 1.31 for
parameters $\omega$ = 0.91, $\theta_{\rm equ}$ = 3.329 mas, and
$\psi$= 40\degr.  A direct comparison of this model with the squared
visibility data is shown in Figure~\ref{fig:v2_bestfit}.

\begin{figure*}
\includegraphics[scale=0.6,angle=0.0]{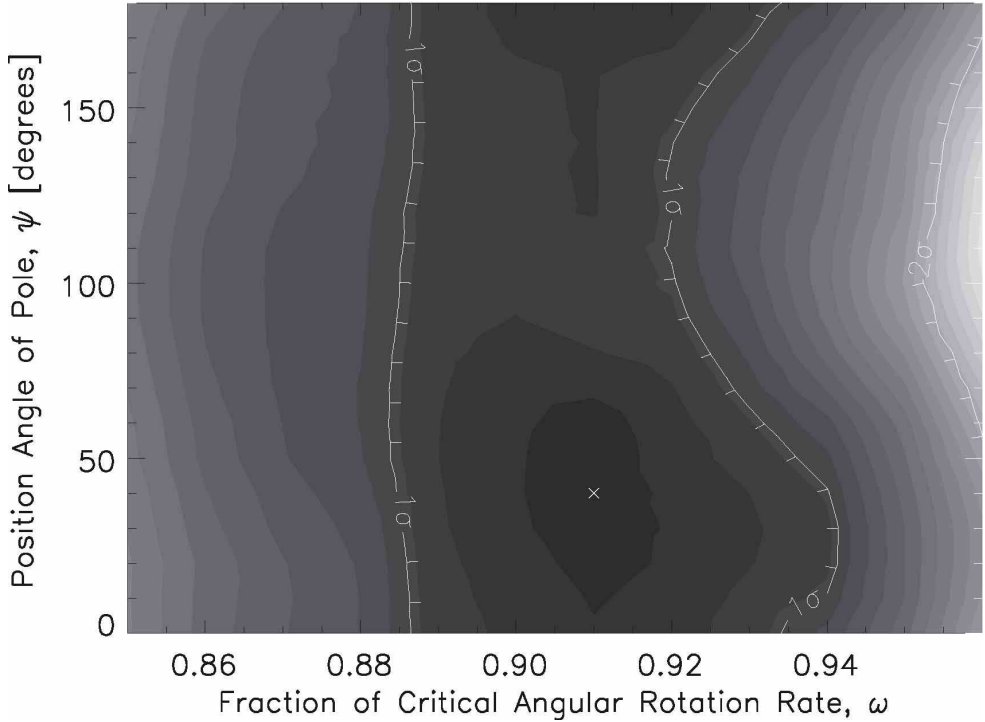}
\caption{A contour plot of $\chi_\nu^2$ in the $\omega - \psi$ plane
for $T^{\rm pole}_{\rm eff}$ = 10250 K and $\log(g)_{\rm pole}$ = 4.1.
The labeled contours denote the lower and upper 1$\sigma$
range, and a 2$\sigma$ contour, from the $F$ test.  The 'x' marks the best fit,
$\chi_\nu^2$=1.31, while the brightest region has a $\chi_\nu^2$ = 3.25 (see
Figure~\ref{fig:chi2_vs_omega}).}
\label{fig:pa_vs_omega}
\end{figure*}


The $F$ test provides a 1$\sigma$ lower limit on $\omega$ at
$\simeq$0.89. For $\omega < 0.89$, the synthetic $V^2$ values are
generally too high across the second lobe because the model is not
sufficiently darkened towards the limb. Correspondingly, the upper
limits on $\omega$ are constrained because the synthetic $V^2$ values
are generally too low across the second lobe, due to very strong
darkening toward the limb for $\omega \gtrsim 0.93$.  In addition, the
upper limit on $\omega$ is a function of $\psi$ because the projected
stellar disk appears sufficiently more elliptical, even at low
inclinations $i \simeq 5\degr$, as the model star rotates faster.  The
data from the nearly orthogonal E2-W1 and E2-S1 baselines constrain
models with $\omega > 0.92$ to limited range of position angles, but
these data provide no constraint on $\psi$ at lower $\omega$ values
where the star is less distorted, $R_{\rm equ}/R_{\rm pole} < 1.24$.

As $\omega$ increases so does the darkening of the limb due to the
increasing larger pole-to-equator effective temperature difference.
As a result, the best fit $\theta_{\rm equ}$ value increases with
$\omega$ because the effective ``limb-darkening'' correction
increases. The best fit values for $\theta_{\rm equ}$ and $\omega$ are
therefore correlated.  To establish this correlation, we estimated the
best fitting $\theta_{\rm equ}$ value for a given $\omega$ without
recomputing the brightness map and Fourier components.  While each
intensity map is constructed for a fixed $\theta_{\rm equ}$ value, we
can approximate the squared visibilities for models with slightly ($<$
0.5\%) larger or smaller $\theta_{\rm equ}$ values as follows.  A small
adjustment to $V^2$ due to a small adjustment in $\theta_{\rm equ}$,
assuming the physical model for the star is not significantly changed
and the model changes relatively slowly with wavelength, is equivalent
to computing $V^2$ at a larger (smaller) wavelength for a larger
(smaller) value of $\theta_{\rm equ}$.  So, for a given projected
baseline, we linearly interpolate (in the log) $V_\lambda^2(u,v)$ at
$\lambda = \lambda_k (\theta_{\rm fit}/\theta_{\rm equ})$, a
wavelength shift of 10 nm or less.  Near the bandpass edges, the
instrument transmission drops to zero so there is no concern about
interpolating outside of the wavelength grid with this scheme. The
$V^2$ normalization, equation (\ref{eqn:norm_quadrature}), must be
scaled by the $(\theta_{\rm fit}/\theta_{\rm equ})^2$ to compensate
for the revised surface area of the star.  After one iteration,
setting $\theta_{\rm equ}=\theta_{\rm fit}$, recomputing the Fourier
map and refitting the data, the best fit $\theta_{\rm equ}$ value is
within 0.25\% of that found with the estimated model $V^2$ values.

\begin{figure*}
\includegraphics[scale=0.6,angle=0.0]{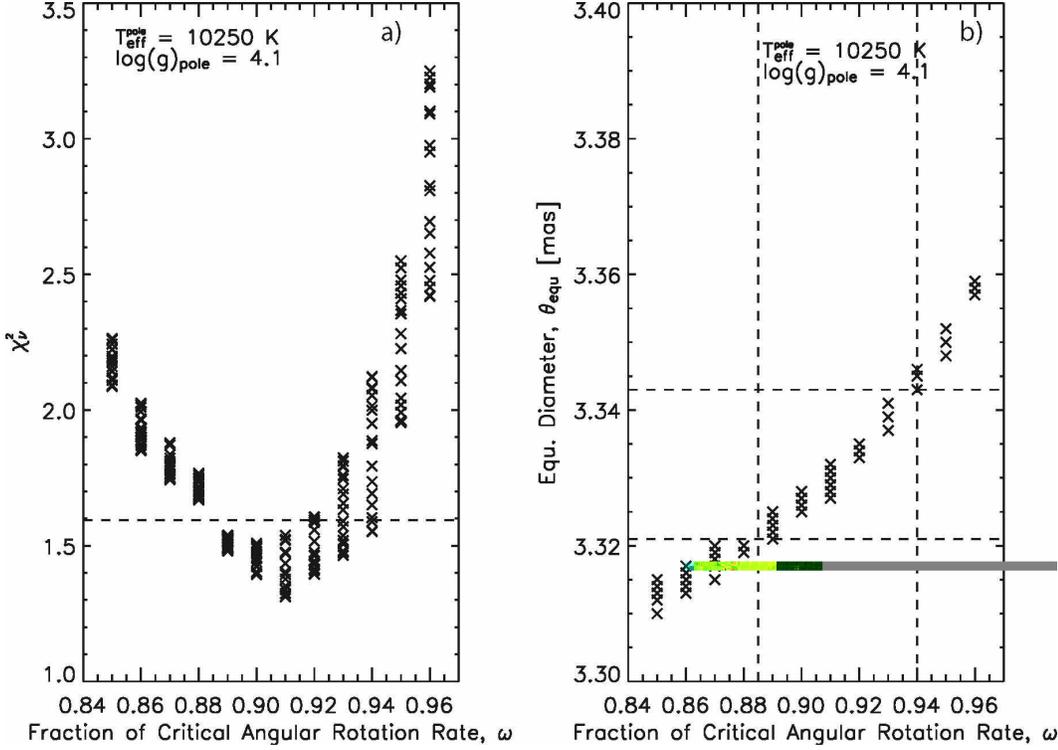}
\caption{Constraints on model parameters from the CHARA/FLUOR data.
a) The reduced chi-square values $\chi_\nu^2$ from the Roche-von
Zeipel model fit to the squared visibility data as a function the
fraction of the critical angular break-up rate,
$\omega=\Omega/\Omega_{\rm crit}$, for fixed values of the polar
effective temperature $T^{\rm pole}_{\rm eff}$ and polar surface
gravity $\log(g)_{\rm pole}$.  The dashed line denotes the 1$\sigma$
confidence region for $\omega$ from the $F$ test for 24 degrees of
freedom relative to the best fit at $\chi_\nu^2$ = 1.31.
For each $\omega$, $\chi_\nu^2$ values are plotted for 18 position angles
$\psi$ (0\degr\ to 170\degr\ in 10\degr\ steps, see
Figure~\ref{fig:pa_vs_omega}).  b) The relationship between the best
fit equatorial angular diameter $\theta_{\rm equ}$ at each $\omega$
for the range of position angles.  The dashed lines provide an
estimate for the 1$\sigma$ range in $\omega$ and the corresponding
range in the equatorial angular diameter.}
\label{fig:chi2_vs_omega}
\end{figure*}

Figure~\ref{fig:chi2_vs_omega}a shows the $\chi_\nu^2$ values from
Figure~\ref{fig:pa_vs_omega} projected on the $\omega$ axis, with a
spread of values for the 18 position angles at each $\omega$ value.
This shows again that for the range $0.89 < \omega < 0.92$ there is no
constraint on the position angle of the pole.  The corresponding best
fit $\theta_{\rm equ}$ values are shown in
Figure~\ref{fig:chi2_vs_omega}b. The  equatorial angular
diameter is constrained to the range $3.32\ {\rm mas} < \theta_{\rm equ}
< 3.34\ {\rm mas}$.  The best fit to the CHARA/FLUOR data is
insensitive to $T^{\rm pole}_{\rm eff}$.  This is because $\Delta\teff$,
which determines the overall darkening, is quite sensitive to
$\omega$, but not $T^{\rm pole}_{\rm eff}$ (see equation
(\ref{eqn:deltateff})).  Thus, we cannot usefully constrain $T^{\rm
pole}_{\rm eff}$ or $\psi$ from the CHARA/FLUOR data.  As for the
surface gravity, varying $\log(g)_{\rm pole}$ over what we consider
the most probable range, 4.1$\pm$0.1, does not significantly effect
the $\chi_\nu^2$ minimum. Models with $\log(g)_{\rm pole}$ values from
3.9 to 4.3 all fall within 1$\sigma$ of the best fit.  The best fit
$\theta_{\rm equ}$ values are essentially independent of $T^{\rm
pole}_{\rm eff}$ between 9800 K and 10450 K and weakly dependent on
$\log(g)_{\rm pole}$ between 3.8 and 4.3; all best fit $\theta_{\rm
equ}$ values fall well within the 1$\sigma$ range established in
Figure~\ref{fig:chi2_vs_omega}.

\subsection{Spectral Energy Distribution: Parameter Grid Search} 

Here we compare our synthetic SEDs to the absolute spectrophotometry
of Vega. Specifically, we compare our models to the data-model
composite
SED\footnote{ftp://ftp.stsci.edu/cdbs/cdbs2/calspec/alpha\_lyr\_stis\_002.fits}
of \cite[]{bg04_vega} which includes {\it International Ultraviolet
Explorer} data from 125.5 to 167.5 nm, {\it HST} Space Telescope
Imaging Spectrograph data from 167.5 to 420 nm, and a specifically
constructed Kurucz model shortward of {\it IUE} and longward 420 nm to
match and replace data corrupted by CCD fringing in this wavelength
region.  To facilitate this comparison, first the synthetic spectra
were convolved to the spectral resolution of the observations
($\lambda/\Delta\lambda =$ 500) and then both the data and convolved
synthetic spectra were binned: 2.0 nm wide bins in the UV (127.5 nm to
327.5 nm, 101 bins) and 2.0 nm bins in the optical and near-IR (330.0
nm to 10080 nm, 340 bins) for a total of 441 spectral bins.

\begin{figure*}
\includegraphics[scale=0.6]{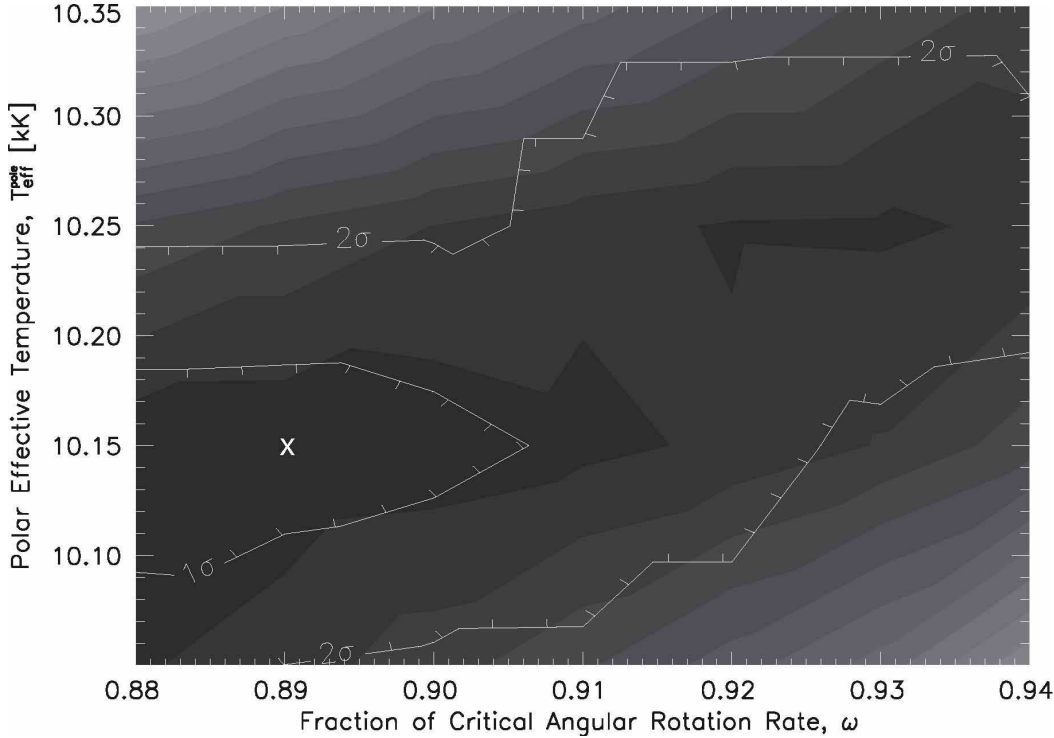}
\caption{A contour plot of $\chi_\nu^2$ for the SED fits in the $\omega -
T^{\rm pole}_{\rm eff}$ plane.  The $\omega$ range is limited to the
1$\sigma$ range from CHARA/FLUOR fits (see
Figure~\ref{fig:chi2_vs_omega}).  The polar surface gravity is fixed
at $\log(g)_{\rm pole}$ = 4.1.  The labeled contours denote the
1$\sigma$ and 2$\sigma$ regions from the $F$ test.  The 'x' marks the
location of the best fit model, $\chi_\nu^2$=8.7.}
\label{fig:sed_chi2_map}
\end{figure*}

Figure~\ref{fig:sed_chi2_map} shows the $\chi_\nu^2$ map in the
$\omega - T^{\rm pole}_{\rm eff}$ plane. These two parameters, apart
from the angular diameter, most sensitively affect the fit to the
observed SED.  There is a clear positive correlation between $\omega$
and $T^{\rm pole}_{\rm eff}$. This makes sense if one considers that a
more rapidly rotating star will be more gravity darkened and require a
hotter pole to compensate for a cooler equator in order to match the
same SED. Following this correlation, it is expected that a continuum
of models from ($\omega =$ 0.89, $T^{\rm pole}_{\rm eff} =$ 10150 K)
to ($\omega =$ 0, $T^{\rm pole}_{\rm eff} =$ 9550 K) will provide a
reasonable fit to the SED since the non-rotating {\tt ATLAS 12} model
of Kurucz fits the observed SED quite well \cite[]{bg04_vega}.
However, we did not consider models with $\omega < 0.88$ in the SED
analysis because such models are a poor match to the CHARA/FLUOR
squared visibility data set as shown above.  In other words, although
the {\tt ATLAS 12} model provides a good fit to the observed SED, it
fails to predict the correct center-to-limb darkening for Vega.

\begin{figure*}
\includegraphics[scale=0.6]{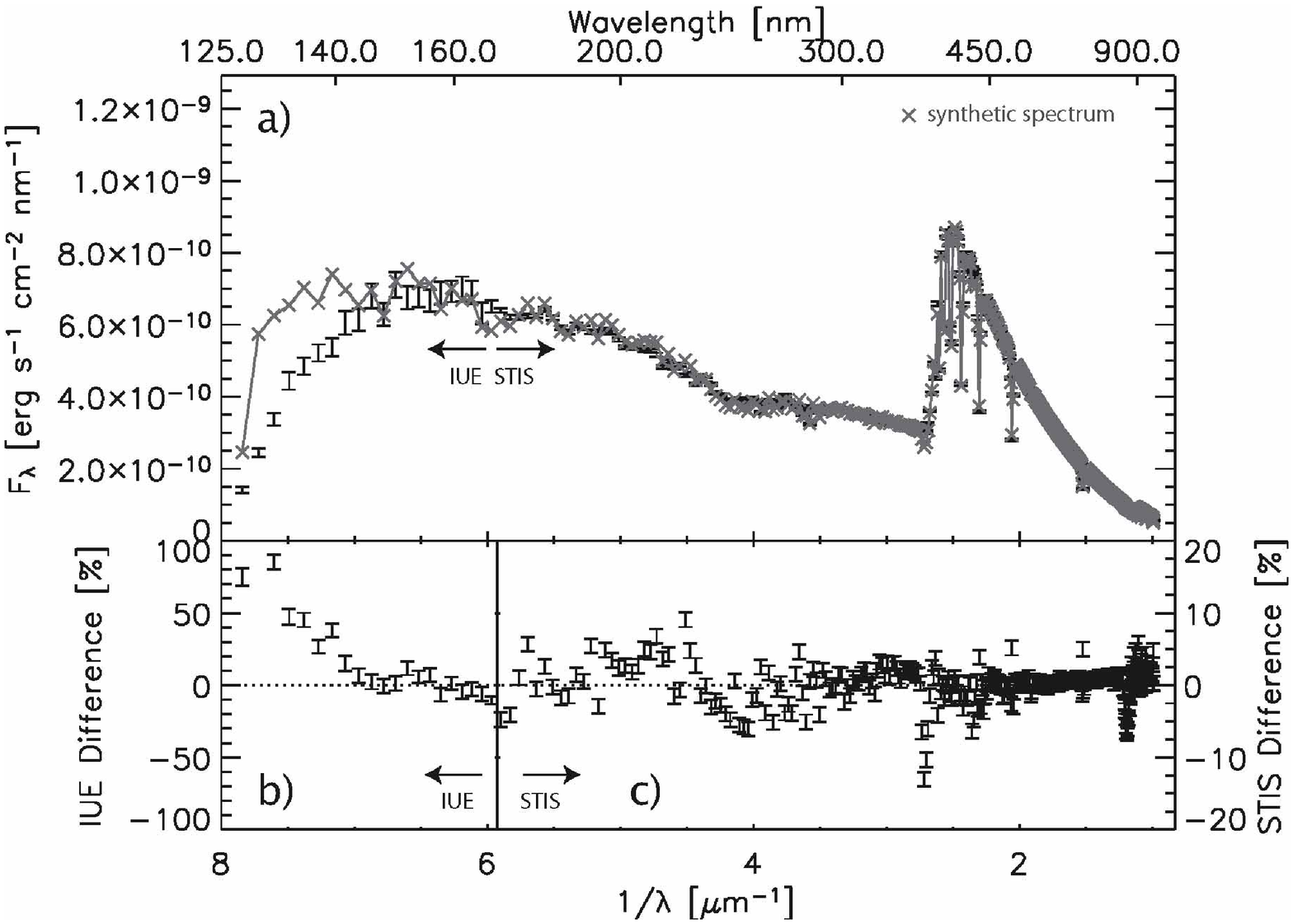}
\caption{a) A comparison between the SED of \cite{bg04_vega}
and our best fitting ($\chi_\nu^2 = 8.7$) rapidly
rotating SED model for Vega: $\omega=0.91$, $T^{\rm pole}_{\rm eff} =$
10150 K and $\log(g)_{\rm pole} =$ 4.10.  The differences between this
model and the data in the b) region at shorter wavelengths observed by
the {\it IUE} and the c) region observed by the {\it HST} 
Space Telescope Imaging Spectrograph at longer wavelengths.
At wavenumbers 1/$\lambda <$ 2.38 $\mu$m$^{-1}$ 
the ``observed'' SED is represented by a closely fitting Kurucz
model spectrum \cite[see][]{bg04_vega}.}
\label{fig:best_fit_sed}
\end{figure*}

The best fit synthetic spectrum is shown in
Figure~\ref{fig:best_fit_sed}.  Considering the complexity of this
synthetic SED relative to a single $\teff$ model, there is generally
good agreement ($\pm$5\%) between our best fit model and the data
longward of 300 nm, apart from larger mismatches at the Paschen and
Balmer edges and in the Balmer lines.  Longward of 140 nm, the model
agrees with the observations to within $\pm$10\%.  At wavelengths
below 140 nm, as measured by the {\it IUE}, the data are up to a
factor of 2 lower than predicted. Our best fit yields
$\chi_\nu^2$=8.7.  The overprediction below 140 nm has only a small
effect on the synthetic integrated flux between 127.5 nm and 10080 nm,
2.79$\times 10^{-5}$ erg cm$^{-2}$ s$^{-1}$, which is within
1.2$\sigma$ of the value derived from an integration of the observed
SED, (2.748$\pm$0.036)$\times 10^{-5}$ erg cm$^{-2}$ s$^{-1}$.  The
equatorial angular diameter derived from this SED fit, $\theta_{\rm
equ} =$ 3.407 mas, differs from the best fit to the CHARA/FLUOR data,
$\theta_{\rm equ} =$ 3.329 mas, by 2.4\%, a value within the
uncertainty of the absolute flux calibration.

\section{Discussion}
\label{discussion}
The best fit stellar parameters, based on the model fits to the
CHARA/FLUOR data and archival spectrophotometry in
\S\ref{2-D_fitting}, are summarized in Table~\ref{tab:fund_params}.
As discussed in \S\ref{1D_models}, the effect of extended K' band
emission in the Vega system, if unaccounted for, is to increase the
apparent angular diameter of Vega slightly, by $\sim$0.3\%.
Correcting for this effect via equation (\ref{eqn:vobs}), the best fit
equatorial diameter is shifted systematically lower by 0.3\% (0.01
mas) to the range $3.31\ {\rm mas} < \theta_{\rm equ} < 3.33\ {\rm
mas}$.  We find all other parameters in Table~\ref{tab:fund_params}
are uneffected by the extended emission within the error bars given.
The best fit range for the fraction of the angular break-up rate,
$0.89 < \omega < 0.92$, sensitive to the amplitude of the second lobe,
is unaffected by the extended emission because the $V^2$ correction is
quite small there, $\Delta V^2 < 0.0003$, relative to the first lobe
where the correction is up to 20 times larger.

\begin{deluxetable*}{llcl}
\tablecolumns{4} 
\tabletypesize{\footnotesize}
\tablecaption{Fundamental Stellar Parameters for Vega} 
\tablewidth{0pt}
\tablehead{
\colhead{Parameter}
&\colhead{Symbol}
&\colhead{Value}
&\colhead{Reference}
}
\startdata              
Fraction of the angular break-up rate      &$\omega$                  &$0.91\pm0.03$              &CHARA/FLUOR $V^2$ fit         \\
Equatorial angular diameter (mas)          &$\theta_{\rm equ}$        &$3.33\pm0.01$              &CHARA/FLUOR $V^2$ fit         \\
Parallax  (mas)                            &$\pi_{\rm hip}$           &$128.93\pm0.55$            &\cite{hipparcos}              \\
Equatorial radius (\rsun)                  &$R_{\rm equ}$             &$2.78\pm0.02$              &Equation (\ref{eqn:requ}) \\
Polar radius      (\rsun)                  &$R_{\rm pole}$            &$2.26\pm0.07$              &Equation (\ref{eqn:rpole}) \\
Pole-to-equator \teff\ difference (K)      &$\Delta\teff$             &$2250^{+400}_{-300}$       &Equation (\ref{eqn:deltateff}) \\
Polar effective temperature (K)            &$T^{\rm pole}_{\rm eff}$  &$10150\pm100$              &Fit to spectrophotometry \cite[]{bg04_vega} \\
Luminosity (\lsun)                         &$L$                       &$37\pm3$                   &Equation (\ref{eqn:lum}) \\ 
Mass   (\msun)                             &$M$                       &$2.3\pm0.2$                &$(L/\lsun)=(M/\msun)^{4.27\pm0.20}$ (from Sirius)\\
Polar surface gravity (cm s$^{-2}$)        &$\log(g)_{\rm pole}$      &$4.1\pm0.1$                &Equation (\ref{eqn:mass}) \\
Equatorial rotation velocity (km s$^{-1}$) &$V_{\rm equ}$             &$270\pm15$                 &Equations (\ref{eqn:Omega}) and (\ref{eqn:vequ}) \\
Projected rotation velocity (km s$^{-1}$)  &$v\,\sin\,i$              &$21.9\pm0.2$               &\cite{hga04} \\
Inclination of rotation axis (degrees)     &$i$                       &$4.7\pm0.3$                &Equation (\ref{eqn:i}) \\
\enddata 
\label{tab:fund_params}
\end{deluxetable*}

\begin{figure*}
\includegraphics[scale=0.6]{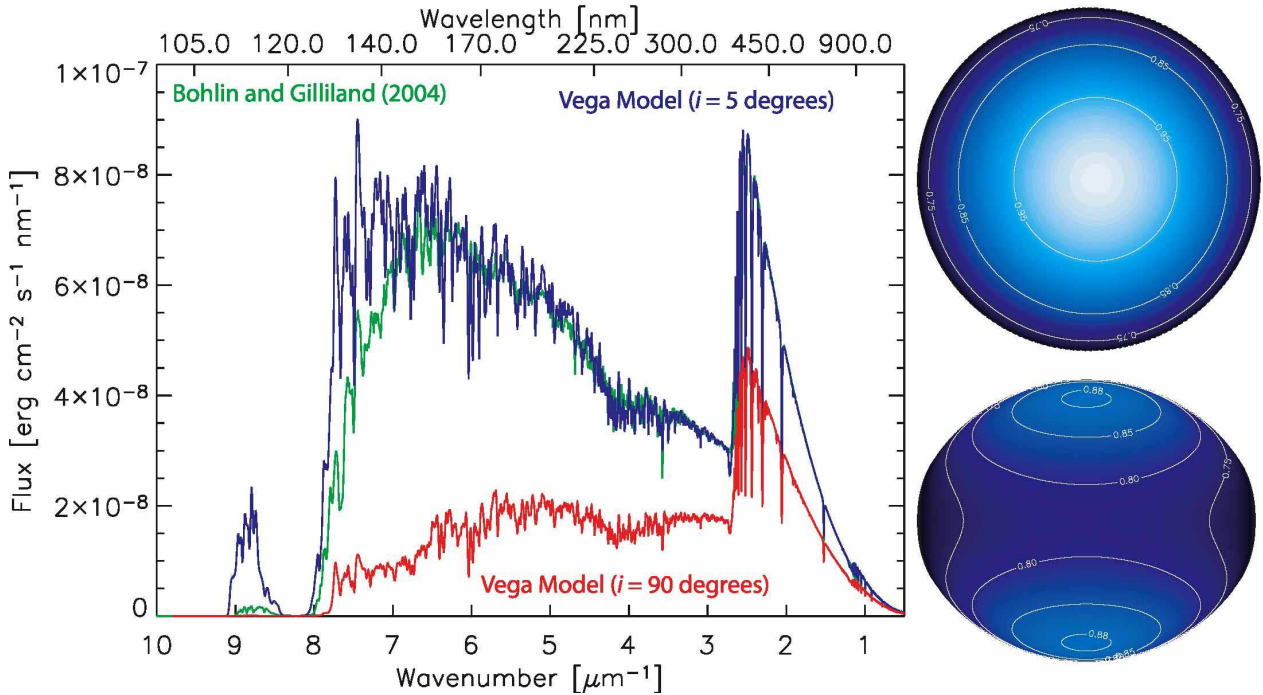}
\caption{Left: A comparison between the SED from 
\cite{bg04_vega} ({\it IUE} and {\it HST} observations supplemented by
a slowly rotating model spectrum both below 127.5 nm and longward of 420
nm) and two rapidly rotating models for Vega's SED, one viewed from an
inclination of 5\degr\ (nearly pole on) and one viewed from an
inclination of 90\degr\ (equator on), from an integration of two
intensity maps via equation (\ref{eqn:flux}) for these inclinations.
Right: A comparison of the best-fit brightness distributions for Vega
with inclinations of 5\degr\ (top) and 90\degr\ (bottom).  For the
equator-on view, the poles appear 10\% fainter than the pole-on view
due to limb darkening.}
\label{fig:sed_comp}
\end{figure*}

One parameter which stands out is our large pole-to-equator effective
temperature difference, $\Delta \teff$ = 2250$^{+400}_{-300}$ K,
relative to previous spectroscopic and spectrophotometric studies of
Vega \cite[]{gha94,hga04} for which $\Delta \teff$ falls into the
range 300 to 400 K.  Our larger $\Delta \teff$ yields a much cooler
equatorial effective temperature, $T_{\rm eff}^{\rm equ}$ =
7900$^{+500}_{-400}$ K, than most recently reported for Vega, 9330 K
\cite[]{hga04}. The amplitude of the second lobe visibility
measurements as observed by CHARA/FLUOR is well fit by strong
darkening toward the limb. In the context of the Roche-von Zeipel
model, such darkening requires a large pole-to-equator \teff\
gradient.  Consequently, we predict that Vega's equator-on SED (that
is viewed as if $i=90$\degr\ and integrated over the visible stellar
disk, see equation (\ref{eqn:flux})) has a significantly lower color
temperature and overall lower flux, particularly in the
mid-ultraviolet where the flux is lower by a factor of 5, as shown in
Figure~\ref{fig:sed_comp}.  A debris disk, aligned with Vega's
equatorial plane as suggested by our nearly pole-on model for the star
and the recent observations of a circular disk in the mid-IR
\cite[]{su05}, should see a significantly less luminous, cooler SED
than we see from the Earth.  In the literature to date, modeling of
the heating, scattering, and emission of Vega's dusty debris disk has
assumed an irradiating SED equal to the pole-on view of Vega \cite[see
e.g.,][]{vega_disk06,su05}. Our synthetic photospheric
equatorial spectrum for Vega is tabulated in Table~\ref{tab:equator_spectrum}.
It should be interesting to investigate how our
predicted equatorial spectrum used in such modeling will affect
conclusions regarding the amount of dust and the grain-size
distribution in the debris disk.

\begin{deluxetable*}{ll}
\tablecolumns{2} 
\tabletypesize{\footnotesize}
\tablecaption{A Model Equatorial Photospheric Spectral Energy Distribution for Vega from 1020.5 \AA\ to 40\micron\ (R=500).
\tablenotemark{a}} 
\tablewidth{0pt}
\tablehead{
\colhead{Wavelength}
&\colhead{Flux ($F_\lambda$)}\tablenotemark{b}\\
\colhead{(\AA)}
&\colhead{(erg cm$^{-2}$ s$^{-1}$ \AA$^{-1}$)}
}
\startdata              
1.020500000000000E+03   &1.68027E+03 \\
1.021000000000000E+03   &1.65680E+03\\
1.021500000000000E+03   &1.62296E+03\\
1.022000000000000E+03   &1.57629E+03\\
1.022500000000000E+03   &1.51370E+03\\
\nodata                 &\nodata\\ 
\enddata 
\tablenotetext{a}{The complete version of this table is in the electronic edition of
the Journal.  The printed edition contains only a sample.}
\tablenotetext{b}{The flux at a distance $d$ from Vega {\it in the equatorial plane}
is the flux from column (2) multiplied by the ratio $(R_{\rm equ}/d)^2$, 
the ratio squared of Vega's equatorial radius to the distance, or $(2.78/d)^2$ when $d$ has the units of solar radii.}
\label{tab:equator_spectrum}
\end{deluxetable*}

Several of Vega's fundamental stellar parameters ($\Delta\teff, V_{\rm
equ}, i$) we derive differ significantly from those derived by
\cite{gha94} and \cite{hga04} from high-dispersion spectroscopy.
Regarding $\Delta\teff$, both spectroscopic studies find $\omega
\simeq 0.5$, while we find $\omega$ = 0.91$\pm$0.03.  These two
$\omega$ values along with the corresponding $T_{\rm eff}^{\rm pole}$
values, 9680 K and 10150 K, in equation (\ref{eqn:deltateff}) yield
$\Delta\teff$ values of 350 K and 2250 K.  The reason the $\omega$
values differ is at least partly linked to inconsistent parameters
used in the spectroscopic studies.  As noted in \cite{hga04}, the
\cite{gha94} study finds a low value for the polar gravity,
$\log(g)_{\rm pole}$ = 3.75, which yields a mass for Vega of only 1.34
\msun\ and an inclination inconsistent with the expected equatorial
velocity. The equatorial velocity of \cite{hga04}, $V_{\rm equ}$ = 160
\kms, is not consistent with their other parameters ($\omega$ = 0.47,
$\log(g)_{\rm pole}$ = 4.0, $R_{\rm equ}$ = 2.73 \rsun, 7.9\degr)
which should yield instead $V_{\rm equ}$ = 113 \kms\ and
$i$=11.1\degr.  Values $V_{\rm equ}$ = 160 \kms\ and $i$ = 7.9\degr\
are recovered if $\omega$ = 0.65, which corresponds to $V_{\rm
equ}/V_{\rm crit}$ = 0.47.  It is possible to confuse $\omega$ with
$V_{\rm equ}/V_{\rm crit}$. The two are not equivalent:
\begin{equation}
\omega = \frac{\Omega}{\Omega_{\rm crit}} \not\equiv \frac{V_{\rm equ}}{V_{\rm crit}} = 
{2\,\cos \biggl[\frac{\displaystyle \pi+\cos^{-1}(\omega)}{\displaystyle 3}\biggr]}.
\label{eqn:little_omega}
\end{equation}
For $\omega$ = 0.65, the corresponding $\Delta\teff$ = 757 K, not 350
K.  Therefore, there appears to be a mismatch between the $V_{\rm
equ}$ and $\Delta\teff$ values used in the most recent spectral
analyses and this suggests the spectral data must be reanalyzed with a
consistent model.  A.~Gulliver (private communication, 2006) confirms
that \cite{hga04} did confuse $\omega$ with $V_{\rm equ}/V_{\rm crit}$
and this group is now reanalyzing Vega's high dispersion spectrum.
Our best fit value for $\omega$, derived from the interferometric data
is appealing because, together with our derived polar effective
temperature, it yields a luminosity consistent with that of slowly
rotating A0 V stars.  A more slowly rotating model for Vega will have
a warmer equator and an overall higher true luminosity too large for
its mass.  Therefore, it seems that less rapidly rotating models for
Vega do not offer an explanation for the apparent over-luminosity with
respect to its spectral type.

Our best fit model, while it provides self-consistent parameters
within the Roche-von Zeipel context, has several discrepancies, most
notably producing too much flux below 140 nm relative to the observed
SED.  The limitations of the LTE metal line blanketing for modeling
Vega in the ultraviolet have recently been explored by
\cite{ggglaph_vega05}.  They find that in the UV the line opacity is
generally systematically too large in LTE because the over-ionization
in non-LTE is neglected.  Our best model flux below 140 nm is already
too large, so a fully non-LTE treatment is not expected to improve
this discrepancy.  The Wien tail of Vega's SED will be the most
sensitive to the warmest colatitudes near the pole.  In our strictly
radiative von Zeipel model, SEDs with $T_{\rm eff}^{\rm pole} <$ 10050
K produce too much flux in the optical and near-IR, so simply lowering
$T_{\rm eff}^{\rm pole}$ will not solve the problem, the temperature
gradient must differ from the $\teff \propto g_{\rm eff}^{0.25}$
relation.  The equatorial effective temperature we derive, 7900 K, may
indicate that Vega's equatorial region is convective.  If so, von
Zeipel's purely radiative gravity darkening exponent, $\beta$ = 0.25,
will not be valid near the equator. A more complex model, where the
gravity darkening transitions from purely radiative near the pole to
partially convective near the equator, may be the next approach to
take.  Such a temperature profile may allow for a cooler $T_{\rm
eff}^{\rm pole}$, reducing the flux discrepancy below 140 nm, while
still matching the observed optical and near-IR fluxes.  Such a
gradient must also improve the match to the Balmer and Paschen edges
and the Balmer lines.

\section{Summary}
\label{summary}
We have demonstrated that a Roche-von Zeipel model atmosphere rotating
at 91$\pm$3\% of the angular break-up rate provides a very good match
to K' band long-baseline interferometric observations of Vega.  These
observations sample the second lobe of Vega's visibility curve and
indicate a limb-darkening correction 2.5 times larger than expected for
a slowly rotating A0 V star.  In the context of the purely radiative
von Zeipel gravity darkening model, the second lobe visibility
measurements imply a $\sim$22\% reduction in the effective temperature
from pole to equator.  The model predicts an equatorial velocity of
270$\pm$15 \kms, which together with the measured $v\sin\,i$ yields an
inclination of $i\simeq 5\degr$, confirming the pole-on model for Vega
suggested by \cite{gray_vega88} to explain Vega's anomalous
luminosity.  Our model predicts a true luminosity for Vega of 37$\pm$3
\lsun, consistent with the mean luminosity of A0 V stars from $W({\rm
H}\gamma)-M_V$ calibration \cite[]{MW85}.  We predict that Vega's
spectral energy distribution viewed from its equatorial plane is
significantly cooler than viewed from its pole. This equatorial
spectrum may significantly impact conclusions derived from models for
Vega's debris disk which have employed Vega's observed polar-view
spectral energy distribution, rather than the equatorial one, which
seems more appropriate given our observations.

\begin{acknowledgments}
We thank G. Romano and P.J. Goldfinger for their assistance with the
operation of FLUOR and CHARA respectively.  F. Schwab kindly provided
advice on two-dimensional FFTs and aliasing.  Thanks to T. Barman for
discussions about numerical cubature and to the entire {\tt PHOENIX}
development team for their support and interest in this work.  Thanks
to referee A. Gulliver for his careful reading and suggestions.
This work was performed in part under contract with the Jet Propulsion
Laboratory (JPL) funded by NASA through the Michelson Fellowship
Program.  JPL is managed for NASA by the California Institute of
Technology.  NOAO is operated by AURA, Inc, under cooperative
agreement with the National Science Foundation.  This research has
been supported by National Science Foundation grants AST­0205297 and
AST­0307562. Additional support has been received from the Research
Program Enhancement program administered by the Vice President for
Research at Georgia State University.  In addition, the CHARA Array is
operated with support from the Keck Foundation and the Packard
Foundation.  This research has made use of NASA's Astrophysics Data
System, and the SIMBAD database, operated at CDS, Strasbourg,
France. Some of the data presented in this paper was obtained from the
Multimission Archive at the Space Telescope Science Institute
(MAST). STScI is operated by the Association of Universities for
Research in Astronomy, Inc., under NASA contract NAS5-26555. Support
for MAST for non-HST data is provided by the NASA Office of Space
Science via grant NAG5-7584 and by other grants and contracts.
\end{acknowledgments}

{\it Facilities:} CHARA (FLUOR)


\end{document}